\documentclass[journal,twoside,web]{ieeecolor}
\usepackage{tmi}
\usepackage{cite}
\usepackage{amsmath,amssymb,amsfonts}
\usepackage{algorithmic}
\usepackage{graphicx}
\usepackage{textcomp}
\usepackage{multirow}
\usepackage{makecell}
\usepackage[bookmarks=true]{hyperref}
\def\BibTeX{{\rm B\kern-.05em{\sc i\kern-.025em b}\kern-.08em
    T\kern-.1667em\lower.7ex\hbox{E}\kern-.125emX}}
\begin{document}
\title{Learning from Ambiguous Labels for Lung Nodule Malignancy Prediction}
\author{Zehui Liao, Yutong Xie, Shishuai Hu, and Yong Xia, \IEEEmembership{Member, IEEE}
\thanks{This work was supported by the National Natural Science Foundation of China under Grants 61771397.
({\em Zehui Liao and Yutong Xie contributed equally to this work.})
({\em Corresponding author: Yong Xia})}
\thanks{The authors are with the National Engineering Laboratory for Integrated Aero-Space-Ground-Ocean Big Data Application Technology, School of Computer Science and Engineering, Northwestern Polytechnical University, Xi'an 710072, China (e-mail: merrical@mail.nwpu.edu.cn; xuyongxie@mail.nwpu.edu.cn; sshu@mail.nwpu.edu.cn; yxia@nwpu.edu.cn).}}

\maketitle

\begin{abstract}
Lung nodule malignancy prediction is an essential step in the early diagnosis of lung cancer. Besides the difficulties commonly discussed, the challenges of this task also come from the ambiguous labels provided by annotators, since deep learning models have in some cases been found to reproduce or amplify human biases.
In this paper, we propose a multi-view `divide-and-rule' (MV-DAR) model to learn from both reliable and ambiguous annotations for lung nodule malignancy prediction on chest CT scans.
According to the consistency and reliability of their annotations, we divide nodules into three sets: a consistent and reliable set (CR-Set), an inconsistent set (IC-Set), and a low reliable set (LR-Set). The nodule in IC-Set is annotated by multiple radiologists inconsistently, and the nodule in LR-Set is annotated by only one radiologist.
Although ambiguous, inconsistent labels tell which label(s) is consistently excluded by all annotators, and the unreliable labels of a cohort of nodules are largely correct from the statistical point of view. Hence, both IC-Set and LR-Set can be used to facilitate the training of MV-DAR.
Our MV-DAR contains three DAR submodels to characterize a lung nodule from three orthographic views. Each DAR consists of a prediction network (Prd-Net), a counterfactual network (CF-Net), and a low reliable network (LR-Net), learning on CR-Set, IC-Set, and LR-Set, respectively. The image representation ability learned by CF-Net and LR-Net is then transferred to Prd-Net by negative-attention module (NA-Module) and consistent-attention module (CA-Module), aiming to boost the prediction ability of Prd-Net.
The MV-DAR model has been evaluated on the LIDC-IDRI dataset and LUNGx dataset. Our results indicate not only the effectiveness of the proposed MV-DAR model in learning from ambiguous labels but also its superiority over present noisy label-learning models in lung nodule malignancy prediction.

\end{abstract}


\section{Introduction}
\label{sec:introduction}
\IEEEPARstart{L}{ung} cancer, which has high morbidity and a low survival rate, is the most common cause of cancer death all over the world \cite{2017Missed}.
According to the latest cancer statistics ~\cite{siegel2021cancer}, an estimated 608,570 Americans will die from cancer in 2021, among which almost one-quarter are due to lung cancer. High lung cancer mortality reflects the large proportion of patients (57\%) diagnosed with metastatic disease, for which the five-year survival is 6\%. However, the five-year survival is 59\% when the primary tumor is small and in the localized stage.
Therefore, the early diagnosis of lung cancer can offer the best chance to cure \cite{aberle2011reduced}. 
The National Lung Screening Trial \cite{aberle2011reduced} shows that there is potential for earlier diagnoses through annual screening with chest computed tomography (CT), which demonstrated a 20\% reduction in lung cancer mortality.

\begin{figure}[t]
\centering
\includegraphics[width=1.0\columnwidth]{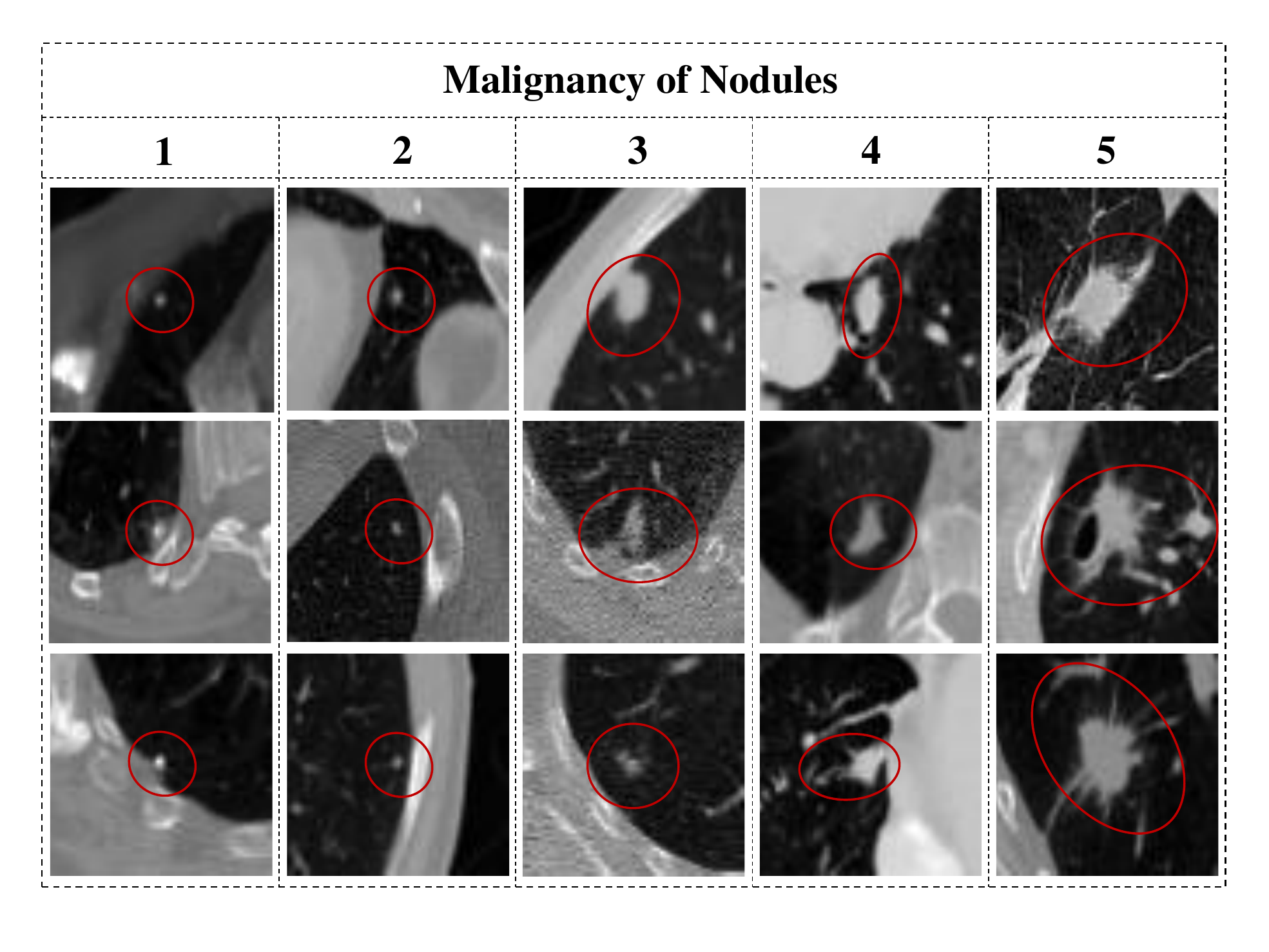} 
\caption{Examples of lung nodules with various malignancies in the axial plane.} 
\label{fig:fig_add1}
\end{figure}

A "spot" less than 3cm on the lung, detected by chest CT, is defined as a lung nodule that could be either benign or malignant. The malignancy of a nodule is usually evaluated with a 5-point scale, from benign to malignant \cite{Ost2003ClinicalPT} (see Fig.~\ref{fig:fig_add1}). 
Since malignant nodules may develop into primary lung tumors or metastatic cancer \cite{Slatore2016WhatIA}, predicting the malignancy of each nodule is a critical step in the early diagnosis of lung cancer. The enormous number of chest CT scans produced globally are currently analyzed almost entirely through visual inspection on a slice-by-slice basis. It requires a high degree of expertise and concentration, and is time-consuming, expensive, and prone to operator bias \cite{Joskowicz2018InterobserverVO}. Automated lung nodule malignancy prediction would enable radiologists to bypass many of these issues and improve the efficiency and accuracy of the diagnosis.

\begin{figure}[t]
\centering
\includegraphics[width=1.0\columnwidth]{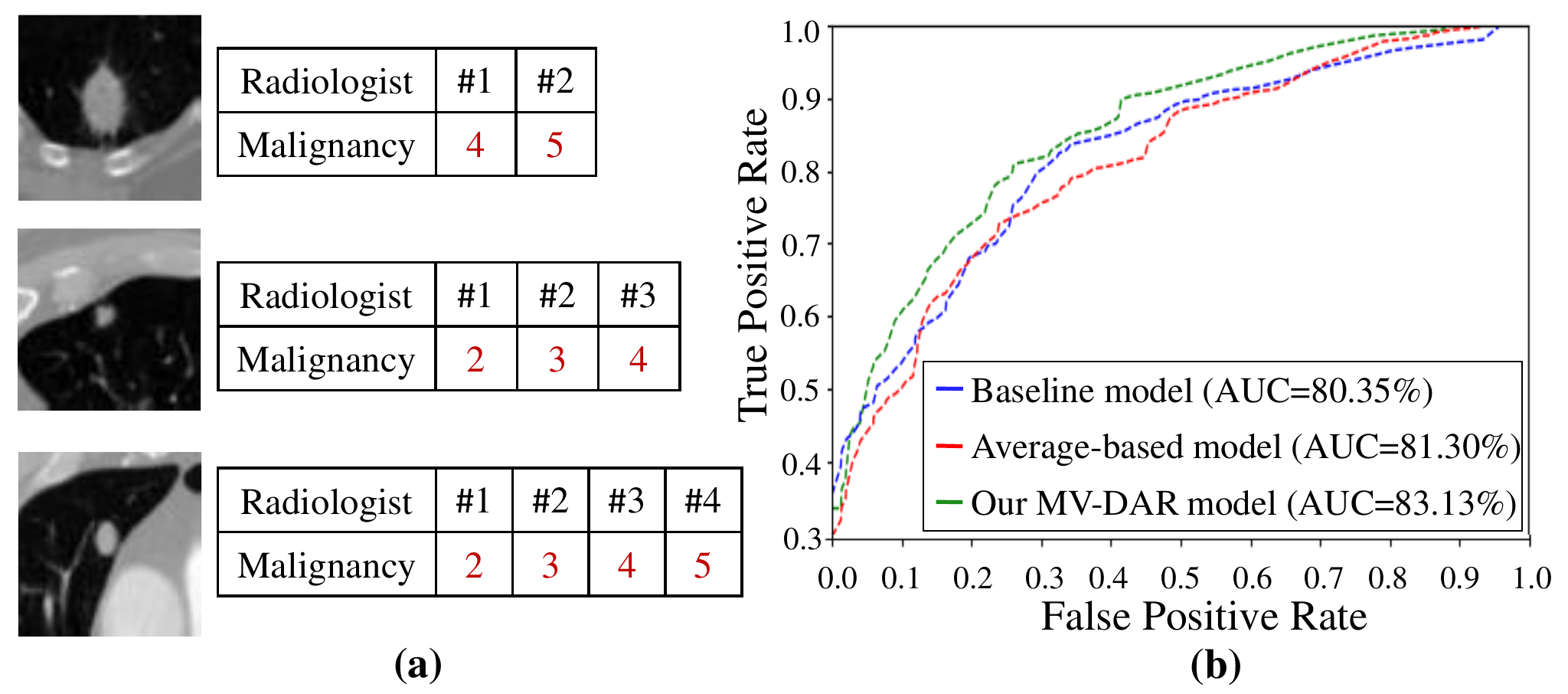} 
\caption{(a) Ambiguous labels of three lung nodules sampled from the LIDC-IDRI dataset. (b) ROC curves of the baseline model (blue), AVE model (red), and proposed MV-DAR model (green). The baseline model was trained with reliably labeled data. The AVE model and our MV-DAR were trained with both reliably labeled and ambiguously labeled data. With the proposed `divide-and-rule' strategy, our MV-DAR substantially outperforms other two models.} 
\label{fig:fig1}
\end{figure}

Recently, deep convolutional neural networks (DCNNs) have achieved impressive performance in lung nodule malignancy prediction\cite{chen2021artificial,gu2021performance,DBLP:journals/mia/XuWGGWBZLY20,DBLP:journals/tmi/LiuDCQH20,al2020procan,afshar20203d,paul2020convolutional,Xie2019KnowledgebasedCD,Wu2019LearningWU,Xie2019SemisupervisedAM,ozdemir20193d, liu2019multi}. However, as a kind of data-driven model, DCNNs have in some cases been found to reproduce or amplify human errors and biases introduced to the training dataset during data acquisition and annotation ~\cite{rich2019lessons}. This problem is particularly true in medical image analysis, where medical image annotation such as determining the malignancy score of a lung nodule is usually prone to divergence, even for experienced medical professionals~\cite{Watadani2013InterobserverVI}. 
Three typical examples of divergent annotations are shown in Fig.~\ref{fig:fig1}(a), where the malignancy scores assigned to the same nodule by different radiologists are inconsistent.
Due to the cost of image acquisition and annotation, we have a small dataset to train a DCNN for lung nodule malignancy prediction, which greatly limits the model performance. Discarding the nodules with inconsistent malignancy scores makes the training dataset even smaller and may lead to even worse performance \cite{ozdemir20193d}.
Therefore, current mainstreams~\cite{jiang2021learning,DBLP:journals/mia/XuWGGWBZLY20,DBLP:journals/tmi/LiuDCQH20,al2020procan,Xie2019KnowledgebasedCD} treat the mean malignancy score as a proxy to converts inconsistent scores into one score. Such an average-annotation strategy is intuitive and easy to implement, but the proxy score may not equal the genuine malignancy score. In other cases, a lung nodule was visually inspected by only one radiologist. Due to the existence of human biases, the malignancy score given by one inspector is unreliable.

Thus, according to the consistency and reliability of annotations, lung nodules can be divided into three subsets, including 
(1) a consistent and reliable set (CR-Set), where each nodule was annotated by multiple radiologists consistently; 
(2) an inconsistent set (IC-Set), where each nodule was annotated by multiple radiologists inconsistently; and 
(3) a low reliable set (LR-Set), where each nodule was annotated by only one radiologist.

The consistent malignancy score in CR-Set is believed to be highly credible and can be used to train and verify a DCNN. Both the proxy malignancy score in IC-Set and unreliable malignancy score in LR-Set may be incorrect. Using those incorrect scores to supervise the training of a DCNN may deteriorate, instead of ameliorating, the model’s performance. However, the information provided by IC-Set and LR-Set is uncertain, but not useless. For a nodule in IC-Set, although we do not know which malignancy score given by radiologists is correct, we can assume that the malignancy score(s) not given by any radiologists is incorrect~\cite{Ishida2017LearningFC}. Therefore, we advocate learning the counter-fact, $i.e.$, the certainly incorrect malignancy scores, on IC-Set.
As for LR-Set, although it is uncertain if the malignancy score of a specific nodule is correct, it is certain from the statistical point of view that the malignancy scores of all nodules in this set are correct with a high probability. Hence, it is still possible to learn the malignancy prediction ability on LR-Set.

In this paper, we propose a “divide-and-rule” (DAR) model to learn from ambiguous labels for lung nodule malignancy prediction on chest CT scans, which can be formulated into a five-class classification problem.
This model can be trained by jointly using the data from CR-Set, IC-Set, and LR-Set while alleviating the impact of inconsistent and unreliable annotations.
First, we construct a malignancy prediction network (Prd-Net) and train it on CR-Set. 
Second, we design a counterfactual learning network (CF-Net) to learn certainly incorrect malignancy scores of each nodule on IC-Set. We devise a negative-attention module (NA-Module) that reversely applies the attention obtained by CF-Net to Prd-Net, aiming to highlight the regions with discriminative information.
Third, we use the nodules in LR-Set to train another malignancy prediction network called LR-Net. We create a consistent-attention module (CA-Module) that utilizes the consistent representations between Prd-Net and LR-Net to push Prd-Net to focus on the discriminative information with high confidence.
Fourth, we merge the outputs of two attention modules with the features from Prd-Net, thus further enhancing the feature representation for accurate malignancy prediction.
Finally, to make use of the context provided by volumetric CT data, we extend our DAR model to the multi-view learning DAR (MV-DAR) model, which has been evaluated on the LIDC-IDRI and LUNGx datasets. 
The receiver operating characteristic (ROC) curves shown in Fig.~\ref{fig:fig1}(b) suggest that the proposed MV-DAR model substantially outperforms the baseline model and average-annotation based (AVE) model.

The contributions of this work include: 
(1) we attempt to address the issue of ambiguous annotations in training data using the divide-and-rule strategy, $i.e.$, dividing the training data with possibly ambiguous labels into three sets and constructing three networks to learn on them, respectively; 
(2) we abandon the idea of using the data with certain or uncertain annotations to train a network without distinction, and instead develop the NA-Module and CA-Module to transfer the representation ability learned on the data with uncertain or unreliable annotations to Prd-Net; and 
(3) the proposed MV-DAR model can learn from ambiguous labels for more accurate lung nodule malignancy prediction, evidenced by its superior performance over other methods on both LIDC-IDRI and LUNGx datasets.

\section{Related Works}

\subsection{Lung nodule malignancy prediction}
Many DCNNs have been proposed in the literature for lung nodule malignancy prediction or benign-malignant nodule classification. 
Xu et al. \cite{DBLP:journals/mia/XuWGGWBZLY20} developed a multi-scale cost-sensitive neural network (MSCS-DeepLN) to overcome the difficulties caused by insufficient training data and class imbalance.
Liu et al. \cite{DBLP:journals/tmi/LiuDCQH20} designed a Multi-Task deep model with Margin Ranking loss (MTMR-Net) to explore the relatedness between lung nodule classification and attribute score regression, which can contribute to the performance gains on both tasks. 
Al-Shabi et al. \cite{al2020procan} proposed a Progressive growing Channel Attentive Non-Local (ProCAN) network with the channel-wise attention for lung nodule classification.
Lei et al. \cite{lei2020shape} developed a soft activation mapping (SAM)-based method for interpretable lung nodule classification.
They further employed the meta-learning scheme and proposed a Meta Ordinal Weighting Network (MOW-Net) and a meta ordinal set (MOS) to explore the ordinal relationship resided in the data itself for lung nodule classification \cite{lei2021meta}.
In our previous work \cite{Xie2019KnowledgebasedCD}, we proposed a multi-view knowledge-based collaborative (MV-KBC) deep model that incorporates the prior knowledge about the high correspondence between a nodule’s malignancy and its heterogeneity shape and voxel values into DCNNs for nodule classification.

Despite their improved performance, these models all use the average-annotation strategy. The average malignancy score, however, may be incorrect when the majority of manual annotations are wrong. Such human errors may be amplified by these models, and hence deteriorate their performance.

\begin{figure*}[t]
\centering
\includegraphics[width=0.97\textwidth]{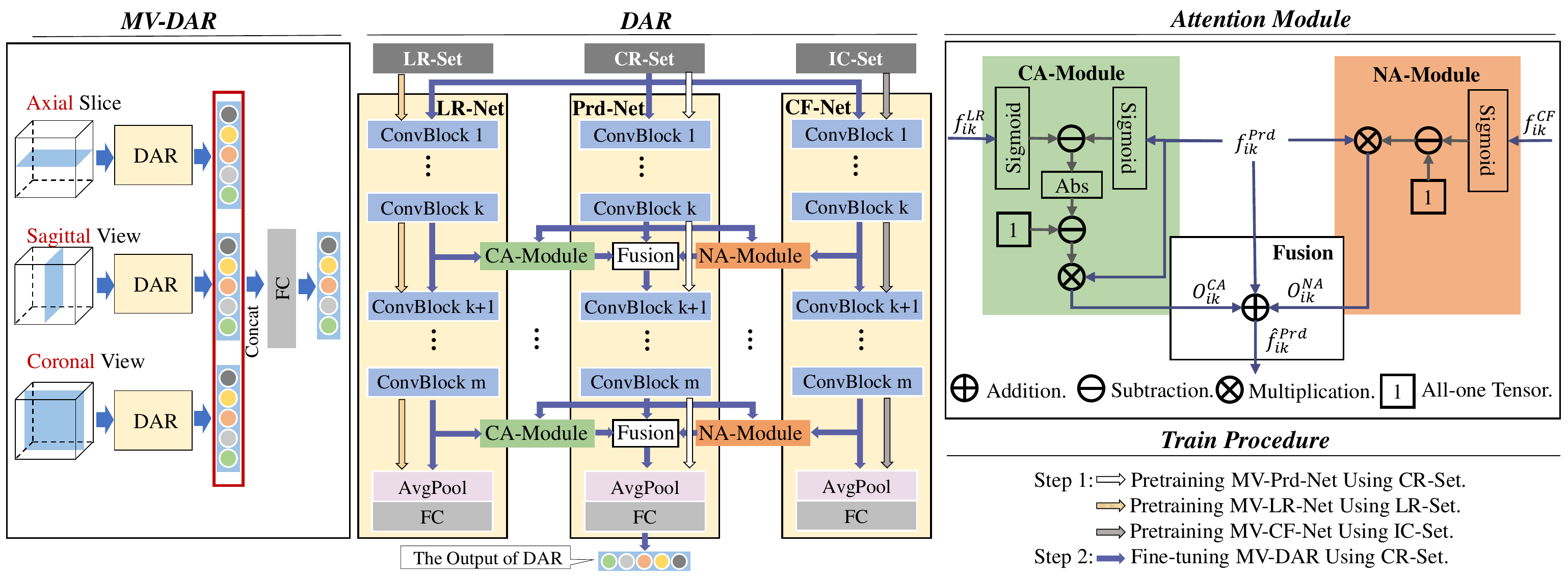}
\caption{The framework of MV-DAR and architecture of each DAR. 'ConvBlock': Convolutional block, 'AvgPool': Global average pooling layer, 'FC': Fully connected layer, and 'Abs': Absolute value. 
Each DAR contains three DCNNs ($i.e.$, Prd-Net, CF-Net, and LR-Net) and two attention modules ($i.e.$, NA-Module and CA-Module) to characterize a nodule from one of three orthographic views ($i.e.$, axial, sagittal, and coronal views). 
Especially, Prd-Net, CF-Net, and LR-Net learn image representations from CR-Set, IC-Set, and LR-Set, respectively. NA-Module and CA-Module transfer the image representation ability learned by CF-Net and LR-Net to Prd-Net, respectively.
}
\label{fig:fig2}
\vspace{-0.2cm}
\end{figure*}

\subsection{Learning from ambiguous labels}
Recently, learning from ambiguously labeled data, including those with an unreliable single label or inconsistent multi-label, has attracted increasing research attention in the computer vision community.
Both noise-model-based and noise-model-free methods have been proposed to learn from noisy labels \cite{Algan2019ImageCW}. The former tries to find the underlying noise structure and then utilizes the structure to avoid the interference caused by noisy labels \cite{Huang2019O2UNetAS}, whereas the latter aims to develop inherently noise-robust strategies using a robust loss function or a regularizer \cite{Wang2019SymmetricCE}.
Although these methods address the unreliability of noise-corrupted labels, they can hardly be extended to address the issue of inconsistent multi-label.

To learn from inconsistent multi-labels, Bayesian probabilistic-based, identification-based, and proxy label-based methods have been proposed. The first stream of methods utilizes the Bayesian probabilistic model to estimate the annotation competence, expertise, and bias of each annotator, and then integrate all candidate labels based on the estimations ~\cite{Simpson2019ABA,Li2019ExploitingWC}. A major limitation of these methods is that they need to use the correspondence between the annotation and its annotator, which in many cases is missing.
To avoid this limitation, identification-based methods~\cite{Dgani2018TrainingAN,Goldberger2017TrainingDN} take the candidate label as a hidden variable and identify it gradually via an iterative expectation–maximization (EM) process. Unfortunately, due to the intrinsic drawback of EM-based optimization, it is impossible to correct false positive labels identified during the process. 
Proxy-label-based methods aim to create a proxy label for each inconsistently labeled sample via computing the mean label~\cite{Xie2019KnowledgebasedCD}, adopting a label sampling strategy~\cite{Jensen2019ImprovingUE}, or using the temporal ensemble of labels~\cite{Yao2020DeepDC}. These methods, however, cannot guarantee the correctness of proxy labels.
In contrast, our MV-DAR model abandons the use of identified labels or proxy labels. Instead, it introduces a novel “divide-and-rule” strategy to learn the image representation ability from ambiguous annotations effectively.

\section{Method}
\label{sec:method}

\subsection{Overview}

The proposed MV-DAR model contains three DAR submodels, each characterizing a nodule from one of three orthographic views. Each DAR consists of three DCNNs ($i.e.$, Prd-Net, CF-Net, and LR-Net) and two attention modules ($i.e.$, NA-Module and CA-Module). Prd-Net, CF-Net, and LR-Net learn image representations on CR-Set, IC-Set, and LR-Set, respectively. Two attention modules transfer the image representation ability learned by CF-Net and LR-Net to Prd-Net, respectively. The pipeline of training DAR can be divided into two stages: pretraining three DCNNs and fine-tuning DAR. The framework of MV-DAR and the architecture of each DAR are illustrated in Fig. \ref{fig:fig2}. We now delve into the details of this model.

\subsection{Pretraining three DCNNs}

\subsubsection{Pretraining Prd-Net on CR-Set}
Since the malignancy score of each nodule in CR-Set is highly credible, we can directly use CR-Set to train Prd-Net for lung nodule malignancy prediction. Let CR-Set be denoted by $I_{CR} = \{(\mathbf{x}_1^{CR},\mathbf{y}_1^{CR}),...,(\mathbf{x}_{N_1}^{CR},\mathbf{y}_{N_1}^{CR})\}$, where $\mathbf{y}_i^{CR}$ is the malignancy score of the $i$-th nodule $\mathbf{x}_i^{CR}$ in CR-Set, and the $N_1$ is the size of CR-Set. The following cross-entropy loss is adopted for Prd-Net

\vspace{-0.3em}
\begin{equation}
L_{Prd}\left ( \widetilde{\mathbf{y}}_{Prd}^{CR},\mathbf{y}^{CR}\right)=-\frac{1}{N_{1}}\sum_{i=1}^{N_{1}}\mathbf{y}_{i}^{CR}\log\left (\widetilde{\mathbf{y}}_{i_{Prd}}^{CR}\right ),
\end{equation}

\noindent where $\widetilde{\mathbf{y}}_{i_{Prd}}^{CR}$ is the predicted malignancy scores of input $\mathbf{x}_{i}^{CR}$.
\vspace{0.4em}

\subsubsection{Pretraining CF-Net on IC-Set}
Each nodule in IC-Set has multiple candidate malignancy scores. To utilize the information embedded in ambiguous labels, we assume that the malignancy scores out of the candidate set are certainly incorrect. Thus, we use CF-Net to learn certainly incorrect malignancy scores ($i.e.$, the counter fact) on IC-Set. Let IC-Set be denoted by $I_{IC}=\{(\mathbf{x}_1^{IC},\mathbf{y}_1^{IC}),…,(\mathbf{x}_{N_2}^{IC},\mathbf{y}_{N_2}^{IC})\}$, where $\mathbf{y}_i^{IC}$ is the set of candidate malignancy scores of the $i$-th nodule $\mathbf{x}_i^{IC}$ in IC-Set and $N_2$ is the size of IC-Set. 

For simplicity, we express $\mathbf{y}_i^{IC}$ as a $Q$-dimensional binary vector, where $1$ occurs at each candidate malignancy score and $Q$ is the number of malignancy levels (i.e. $Q=5$ for this study). The loss function of CF-Net is defined as follows

\vspace{-0.3em}
\begin{equation} 
L_{CF}\left ( \widetilde{\mathbf{y}}_{CF}^{IC},\mathbf{y}^{IC}\right)=-\frac{1}{N_{2}}\sum_{i=1}^{N_{2}}\left ( \textbf{1}-\mathbf{y}_{i}^{IC}\right )\log\left ( \widetilde{\mathbf{y}}_{i_{CF}}^{IC}\right ),
\end{equation}

\noindent where $\widetilde{\mathbf{y}}_{i_{CF}}^{IC}$ is the malignancy scores of $\mathbf{x}_{i}^{IC}$ predicted by CF-Net, and \textbf{1} is a $Q$-dimensional all-one vector.
\vspace{0.4em}

\subsubsection{Pretraining LR-Net on LR-Set}
Since the majority part of LR-Set is annotated correctly, we can still use this dataset to train LR-Net for nodule representation learning. Let LR-Set be denoted by $I_{LR}=\{(\mathbf{x}_1^{LR},\mathbf{y}_1^{LR}),…,(\mathbf{x}_{N_3}^{LR},\mathbf{y}_{N_3}^{LR})\}$, where $\mathbf{y}_i^{LR}$ is the malignancy scores of the $i$-th nodule $\mathbf{x}_i^{LR}$ in LR-Set and $N_3$ is the size of LR-Set.
We also adopt the cross-entropy loss for LR-Net, shown as follows

\vspace{-0.3em}
\begin{equation}
L_{LR}\left ( \widetilde{\mathbf{y}}_{LR}^{LR},\mathbf{y}^{LR}\right)=-\frac{1}{N_{3}}\sum_{i=1}^{N_{3}}\mathbf{y}_{i}^{LR} \log\left ( \widetilde{\mathbf{y}}_{i_{LR}}^{LR}\right ), 
\end{equation}

\noindent where $\widetilde{\mathbf{y}}_{i_{LR}}^{LR}$ is the predicted malignancy scores of $\mathbf{x}_{i}^{LR}$.

\subsection{Fine-tuning DAR on CR-Set}
After pretraining Prd-Net, IC-Net, and LR-Net, we further fine-tune the DAR submodel on CR-Set, with the emphasis on training two attention modules to transfer the image representation ability learned by CF-Net and LR-Net to Prd-Net.

\subsubsection{NA-Module} 
NA-Module is used to bridge CF-Net and Prd-Net by transferring the feature maps obtained by the $k$-th to $m$-th convolutional blocks of CF-Net to the corresponding blocks of Prd-Net, where $m$ is the total number of convolutional blocks. Given the $i$-th input $\mathbf{x}_i^{CR}$, we denote the outputs of the $k$-th convolutional block in Prd-Net and CF-Net as $\mathbf{f}_{ik}^{Prd}\in R^{C\times H\times W}$ and $\mathbf{f}_{ik}^{CF}\in R^{C\times H\times W}$, respectively, where $C$, $H$, and $W$ denote the channel number, height, and width of the feature maps. The output of NA-Module is

\begin{equation}
\mathbf{O}_{ik}^{NA}=(\textbf{1}-\sigma \left ( \mathbf{f}_{ik}^{CF}\right )) \times \mathbf{f}_{ik}^{Prd},
\end{equation}

\noindent where \textbf{1} is a all-one vector with the same size to $\mathbf{f}_{ik}^{CF}$, and $\sigma \left (\cdot \right )$ is the sigmoid function. Since CF-Net learns from incorrect labels, the $\mathbf{f}_{ik}^{CF}$ indicates the mis-leading information. Thus, we first invert the normalized $\mathbf{f}_{ik}^{CF}$ and then apply it to $\mathbf{f}_{ik}^{Prd}$ by performing the element-wise multiplication to obtain the attention map that highlights the regions related to the correct label.

\subsubsection{CA-Module} 
Similarly, CA-Module is used to bridge LR-Net and Prd-Net by transferring the feature maps obtained by the $k$-th to $m$-th convolutional blocks of LR-Net to the corresponding blocks of Prd-Net. Given the $i$-th input $\mathbf{x}_i^{CR}$, we denote the output of the $k$-th convolutional block in LR-Net as $\mathbf{f}_{ik}^{LR}\in R^{C\times H\times W}$. 
Specially, we introduce the consistency between $\mathbf{f}_{ik}^{LR}$ and $\mathbf{f}_{ik}^{Prd}$ as a constraint to Prd-Net, forcing Prd-Net to focus on high-confidence regions, which may contain rich discriminative information. The consistent metric $\mathbf{CM}\in R^{C\times H\times W}$ is calculated as 

\begin{equation}
\mathbf{CM}=\textbf{1}-\left | \sigma \left ( \mathbf{f}_{ik}^{LR}\right )-\sigma \left ( \mathbf{f}_{ik}^{Prd}\right )\right |.
\end{equation}

Next, we perform the element-wise multiplication between $\mathbf{CM}$ and $\mathbf{f}_{ik}^{Prd}$ to obtain the attention map $\mathbf{O}_{ik}^{CA}$.

Finally, the attention maps $\mathbf{O}_{ik}^{NA}$ and $\mathbf{O}_{ik}^{CA}$ are fused with the original classification features $\mathbf{f}_{ik}^{Prd}$ via the element-wise summation to obtain the augmented features $\hat{\mathbf{f}}_{ik}^{Prd}$, shown as follows:

\begin{equation}
\hat{\mathbf{f}}_{ik}^{Prd}=\mathbf{f}_{ik}^{Prd}+\mathbf{O}_{ik}^{NA}+\mathbf{O}_{ik}^{CA}
\end{equation}

Subsequently, the augmented $\hat{\mathbf{f}}_{ik}^{Prd}$ will replace $\mathbf{f}_{ik}^{Prd}$.

\subsubsection{Optimization Loss}
Given an input data $\mathbf{x}^{CR}$, the loss function of the proposed DAR submodel is defined as follows

\begin{equation}
\begin{aligned}
L_{DAR}=&L_{Prd}\left ( \widetilde{\mathbf{y}}_{Prd}^{CR},\mathbf{y}^{CR}\right )\\
&+\mu \times L_{CF}\left ( \widetilde{\mathbf{y}}_{CF}^{CR},\mathbf{y}^{CR}\right )\\
&+\delta \times L_{LR}\left ( \widetilde{\mathbf{y}}_{LR}^{CR},\mathbf{y}^{CR}\right ),
\end{aligned}
\end{equation}
where $\widetilde{\mathbf{y}}_{Prd}^{CR}$, $\widetilde{\mathbf{y}}_{CF}^{CR}$ and $\widetilde{\mathbf{y}}_{LR}^{CR}$ are the predictions made by Prd-Net, CF-Net, and LR-Net, respectively, and $\mu$ and $\delta$ are weighting factors.

\subsection{Extending to MV-DAR}
We extend the proposed DAR submodel to MV-DAR, a multi-view learning model, in two steps: (1) extracting three 2D slices on the axial, sagittal, and coronal planes from each 3D lung nodule, (2) constructing and training the MV-DAR model for nodule malignancy prediction. 

We resample all chest CT scans to an unified voxel size of $1.0\times 1.0\times 1.0$ $mm^{3}$. For each lung nodule, we first crop a $64\times64\times64$ cube that is centered on the nodule location such that the nodule is always contained completely in the cube, and then extract 2D patches on the axial, sagittal, and coronal planes, respectively. Thus, we obtain three views of patches for each nodule. Then, all patches are resized to $224\times224$.

The patches extracted from three views are used to train the MV-DAR model, which contains three DAR submodels that characterize the nodule from axial, sagittal, and coronal views, respectively. Given the $i$-th reliably labeled input patch triplet $\left \{\mathbf{x}_{i}^{CR^{v1}},\mathbf{x}_{i}^{CR^{v2}},\mathbf{x}_{i}^{CR^{v3}}\right \}$, we denote the output of Prd-Net in each DAR as $\widetilde{\mathbf{y}}_{i_{Prd}}^{CR(\#)}$, where $\# \in \left \{v1,v2,v3\right \}$. Thus, the output $\mathbf{P}_{i}$ of MV-DAR can be calculated as

\begin{equation}
    \mathbf{P}_{i} = FC\left ( concat\left ( \widetilde{\mathbf{y}}_{i_{Prd}}^{CR^{v1}}, \widetilde{\mathbf{y}}_{i_{Prd}}^{CR^{v2}}, \widetilde{\mathbf{y}}_{i_{Prd}}^{CR^{v3}} \right ) \right )
\end{equation} 
where $FC$ refers to a fully connected (FC) layer with $Q$ neurons, and $concat$ denotes the concatenation operation.

\section{Experiments}
\subsection{Dataset and Evaluation Metrics}
The largest public lung nodule dataset LIDC-IDRI \cite{Armato2011TheLI,Clark2013TheCI} was used for this study. It contains 2568 lung nodules from 1018 chest CT scans obtained from seven institutions. The nodule diameters range from 3mm to 30mm. Each suspicious lesion is categorized as a non-nodule,  a nodule $<$3 mm, or a nodule $\geqslant$3 mm diameter in the long axis. For this study, we only considered nodules $\geqslant$3 mm in diameter, since nodules $<$3mm were not considered to be clinically relevant by current screening protocols \cite{han2015texture,dhara2016combination,hussein2017tumornet,shen2017multi,setio2016pulmonary,Xie2019KnowledgebasedCD}. The malignancy score of each nodule ranges from one (highly unlikely for cancer) to five (highly likely for cancer) and was annotated by one to four radiologists. Considering the consistency and reliability of these annotations, we divided these nodules into three subsets: a CR-Set with 394 consistently annotated nodules, an IC-Set with 1048 inconsistently annotated nodules, and an LR-Set with 766 unreliably annotated (by one radiologist) nodules. The malignancy distribution of the nodules in each subset is summarized in Table~\ref{tab:table1}. In addition, since the nodules in IC-Set may correspond to multiple candidate category annotations, we count the nodules in each possible category separately.

We evaluated the obtained predictions using the mean and standard deviation of four performance metrics, including the accuracy, recall, area under ROC curve (AUC), and F1-score. Accuracy represents the proportion of samples that are correctly classified by the model, recall shows the proportion of positive samples that are correctly identified, the F1-score is an indicator that comprehensively considers precision and recall, and AUC reflects the expected generalization performance of the model. Since we treat nodule malignancy prediction as a five-class classification problem, the recall, F1-score, and AUC values reported in this paper measure the average performance metrics per class.

\begin{table}[t]
\caption{Malignancy distribution of 2568 lung nodules in three sets. Note that a nodule in IC-Set has multiple malignancy scores.}
\label{tab:table1}
\footnotesize
\centering
\renewcommand{\arraystretch}{1.25}
\begin{tabular}{ccccccc}
\hline
\hline
\multirow{2}{*}{Datasets} & \multicolumn{5}{c}{Malignancy} & \multirow{2}{*}{Number} \\ \cline{2-6}
                          & 1    & 2    & 3    & 4   & 5   &                         \\ \hline
CR-Set                    & 149  & 63   & 143  & 10  & 29  & 394                     \\ 
LR-Set                    & 120  & 250  & 332  & 50  & 14  & 766                     \\ 
IC-Set                    & 280    & 890    & 1144    & 658   & 356   & 1408                    \\ \hline \hline
\end{tabular}
\end{table}

\subsection{Implementation Details}
Since only CR-Set can provide the nodules with highly credible malignancy scores, we evaluated the proposed MV-DAR model on CR-Set using the five-fold cross-validation. In each trial, we randomly chose 10\% of training data to form a validation set, which was used to monitor the training process and prevent overfitting. Note that both IC-Set and LR-Set were merely used as external training data. Considering the impact of randomness on the performance, we repeated the cross-validation five times and reported the mean and standard deviation of each performance metric.

To reduce the number of parameters and computational burden, we chose the EfficientNet-B0~\cite{Tan2019EfficientNetRM}, which was pretrained on the ImageNet dataset~\cite{Deng2009ImageNetAL}, as the backbone of Prd-Net, CF-Net, and LR-Net. As a lightweight model, EfficientNet-B0 contains 16 convolutional blocks ($i.e.$, MBConv). Thus the parameter $m$ in NA-Module and CA-Module equals to 16. The hyper-parameter $k$ was empirically set to 11. To adapt EfficientNet-B0 to our classification task, we removed its last FC layer, and then added a new FC layer with $Q$ neurons. We set the activation function in the last layer to softmax in Prd-Net and LR-Net, and sigmoid in CF-Net.

To enlarge the training dataset, we employed the online data augmentation, which includes randomly horizontal and vertical flips and random rotation from -90 to 90 degrees. We adopted the Adam algorithm~\cite{Kingma2015AdamAM} with a batch size of 32 to optimize MV-DAR. The learning rate was initialized to 0.0001 in the pretraining stage and 0.00005 in the fine-tuning stage, and was reduced by the “poly” policy~\cite{Chen2018EncoderDecoderWA}. The maximum epoch number was set to 100. We set the hyper-parameters in the loss of DAR as $\mu$=0.5 and $\delta$=0.5. 

\begin{table*}[t]
\caption{Performance of our MV-DAR model and five ambiguous label-based models on the LIDC-IDRI dataset. The p-values of all paired t-tests are smaller than the significance level of 0.05, which suggests that the performance gain is statistically significant.}
\label{tab:table2}
\footnotesize
\centering
\renewcommand{\arraystretch}{1.25}
\begin{tabular}{cccccccc}
\hline \hline
\multirow{2}{*}{Methods} 			& \multirow{2}{*}{Backbone} 	& \multirow{2}{*}{\begin{tabular}[c]{@{}c@{}}Params\\($\times 10^6$)\end{tabular}} 	& \multicolumn{4}{c}{Results (\%) (Mean$\pm$standard deviation)}        & \multirow{2}{*}{\begin{tabular}[c]{@{}c@{}}p-value\\(Accuracy)\end{tabular}}               \\ \cline{4-7} 
									&                           	&                         	& Accuracy     				& AUC          							& Recall       				& F1-score       &                                      \\ \hline
MV-AVE		             			& EfficientNet-b0              	& 15.9            			& 65.79$\pm$0.18    		& 81.30$\pm$0.09          	          	& 48.67$\pm$0.31          	& 48.98$\pm$0.43  & 1.15$\times 10^{-8}$                  \\
MV-NAL~\cite{Goldberger2017TrainingDN}&EfficientNet-b0              & 15.9            			& 67.82$\pm$0.05    		& 70.44$\pm$0.07          	          	& 38.15$\pm$0.04          	& 36.55$\pm$0.05  & 5.10$\times 10^{-16}$                  \\
MV-NLNL~\cite{Kim2019NLNLNL}		& EfficientNet-b0              	& 15.9            			& 68.95$\pm$0.04    		& 79.62$\pm$0.05          	          	& 45.49$\pm$0.08          	& 45.74$\pm$0.07    & 1.89$\times 10^{-16}$                \\
MV-LS~\cite{Jensen2019ImprovingUE}  & EfficientNet-b0              	& 15.9            			& 61.61$\pm$0.18    		& \textbf{83.28}$\pm$0.11 	          	& 44.69$\pm$0.29          	& 41.53$\pm$0.16  & 2.08$\times 10^{-9}$                  \\
MV-D$^2$CNN~\cite{Yao2020DeepDC}    & EfficientNet-b0              	& 15.9            			& 71.72$\pm$0.18    		& 79.15$\pm$0.32          	 			& 54.20$\pm$0.27          	& 54.88$\pm$0.23    & 1.60$\times 10^{-6}$                \\
MV-D$^2$CNN*~\cite{Yao2020DeepDC}    & EfficientNet-b4              	& 57.0            			& 72.55$\pm$0.05 			& 83.27$\pm$0.07 			 	& 56.47$\pm$0.08 			& 56.31$\pm$0.08   & 5.07$\times 10^{-12}$                \\
MV-DAR             					& EfficientNet-b0              	& 47.7            			    & \textbf{74.57}$\pm$0.04 	& 83.13$\pm$0.06   			   			& \textbf{58.29}$\pm$0.15 	& \textbf{56.36}$\pm$0.11  & -                      \\ \hline \hline
\end{tabular}
\end{table*}

\subsection{Experimental Results}
\subsubsection{Comparison to ambiguous label-based models}
We compared the proposed MV-DAR model with five ambiguous label learning models, including the AVE model, noise adaptation layer-based (NAL) model~\cite{Goldberger2017TrainingDN}, negative learning (NLNL) model~\cite{Kim2019NLNLNL}, label sampling (LS) model~\cite{Jensen2019ImprovingUE}, and deep discriminative CNN (D$^{2}$CNN) model~\cite{Yao2020DeepDC}. 
The NAL model treats the noise label as a latent random variable and models the noise process by a communication channel. 
The AVE model uses the average of candidate scores as the proxy label, and then trains a classification network using proxy labels. 
The LS model randomly selects the proxy label from candidate scores to train a classification network. 
By contrast, the NLNL model randomly selects the proxy label from non-candidate scores. 
To enhance the confidence level of proxy labels, D$^{2}$CNN assembles the network predictions of the different epochs as proxy supervision for the subsequent epoch.

For a fair comparison, we used the same multi-view learning framework but replaced our DAR with AVE, NAL, NLNL, LS, and D$^{2}$CNN, respectively. Besides, we adopted the same backbone architecture in AVE, NAL, NLNL, LS, and D$^{2}$CNN, and used the same training settings, including the optimizer, learning rate, and batch size.

The results of nodule malignancy prediction produced by these models were given in Table~\ref{tab:table2}. It shows that our MV-DAR model achieves the highest accuracy, recall, and F1-score, and the third-highest AUC. 
In particular, the highest recall ($i.e.$ lowest false negative rate) of MV-DAR indicates that our model is more suitable for lung nodule screening than other models. 
Moreover, the superior performance of MV-DAR also corroborates that our learning strategy avoiding using inaccurate proxy labels or identified labels has a stronger ability to learn the rich potential discriminative information from ambiguous labels than the identification-based model and four proxy-label-based models. 

Moreover, each DAR contains three sub-networks, while the other five comparison models have only one network. 
The prediction performance could be related to the number of model parameters. 
For a fair comparison, we strengthened the most competitive model D$^{2}$CNN by scaling its backbone to EfficientNet-B4, which has nearly four times as many parameters as EfficientNet-B0. Compared to the stronger D$^{2}$CNN* with the EfficientNet-B4 backbone, our DAR with the EfficientNet-B0 backbone has fewer parameters (47.7M vs. 57.0M) but achieves higher accuracy (74.57±0.04\% vs. 72.55±0.05\%).

\begin{table}[t]
\caption{Performance of MV-DAR model under different settings.}
\label{tab:table3}
\footnotesize
\centering
\renewcommand{\arraystretch}{1.25} 
\setlength{\arraycolsep}{0.5pt}{
\begin{tabular}{ccccccc}

\hline \hline
\multirow{2}{*}{Methods} & \multicolumn{4}{c}{Results (\%) (Mean$\pm$standard deviation)}                 \\ \cline{2-5} 
                         & Accuracy   & AUC          & Recall     & F1-score   \\ \hline
SV-Prd                   & 68.88$\pm$0.08 & 81.56$\pm$0.14  & 51.02$\pm$0.15 & 50.15$\pm$0.16 \\
MV-Prd                   & 71.27$\pm$0.07 & 80.35$\pm$0.16  & 53.06$\pm$0.20 & 52.49$\pm$0.19 \\
MV-Prd-LR                & 72.90$\pm$0.05 & 82.87$\pm$0.12  & 55.63$\pm$0.09 & 54.45$\pm$0.11 \\ 
MV-Prd-CF                & 74.48$\pm$0.12 & 82.25$\pm$0.06  & 56.88$\pm$0.22 & 55.30$\pm$0.14 \\ 
MV-Fusion                & 70.84$\pm$0.06 & 82.40$\pm$0.06  & 55.03$\pm$0.12 & 55.37$\pm$0.10 \\
MV-DAR                   & \textbf{74.57}$\pm$0.04 & \textbf{83.13}$\pm$0.06  & \textbf{58.29}$\pm$0.15 &\textbf{56.36}$\pm$0.12 \\ \hline \hline
\end{tabular}}
\end{table}

\subsubsection{Ablation study}
In the proposed MV-DAR model, NA-Module and CA-Module are used to transfer the image representation ability learned by CF-Net and LR-Net to Prd-Net. To demonstrate the contributions of both modules, we conducted ablation studies via constructing the MV-Prd model, MV-Prd-LR model, MV-Prd-CF model, and MV-Fusion model. 
In MV-Prd, each DAR contains only Prd-Net (i.e. LR-Net and CF-Net were removed). 
In MV-Prd-LR, each DAR includes only LR-Net and Prd-Net (i.e. CF-Net was removed). 
In MV-Prd-CF, each DAR includes only CF-Net and Prd-Net (i.e. LR-Net was removed). 
In MV-Fusion, we removed the CA-Module and NA-Module while keeping LR-Net and CF-Net, and fused the decisions made by three sub-networks.
Besides NA-Module and CA-Module, multi-view learning also plays a pivotal role in the MV-DAR model. Thus, we also compared MV-Prd with the single-view Prd (SV-Prd) that characterizes a nodule only in the coronal plane.

The performance of our MV-DAR model and five variants was given in Table~\ref{tab:table3}. It shows that 
(1) embedding Prd-Net into the multi-view architecture can capture more nodule information and achieve better classification performance than the single-view architecture; 
(2) the performance gain of MV-Prd-LR model over MV-Prd and the gain of MV-Prd-CF over MV-Prd demonstrate the effectiveness of LR-Net and CF-Net; 
(3) our MV-DAR model achieves higher accuracy, AUC, recall, and F1-score than MV-Prd-LR and MV-Prd-CF, which indicates that jointly using the LR-Net and CF-Net is superior over using each of them alone; and 
(4) our MV-DAR model beats MV-Fusion, which proves the effectiveness of NA-Module and CA-Module.

To further validate our MV-DAR, we randomly selected ten single view ROIs of lung nodule in LIDC-IDRI dataset and visualized the corresponding patches and learned feature maps in Fig.~\ref{fig:fig4}. In each column, the first is the nodule patch, and the rest four images show the normalized sum of ${\mathbf{f}}_{ik}^{Prd}$, ${\mathbf{f}}_{ik}^{LR}$, ${\mathbf{f}}_{ik}^{CF}$, and $\hat{\mathbf{f}}_{ik}^{Prd}$, where $k=11$.
It shows that 
(1) ${\mathbf{f}}_{ik}^{Prd}$ learned from CR-Set pays more attention to the nodule itself than the ${\mathbf{f}}_{ik}^{LR}$
learned from LR-Set. 
(2) the feature maps learned from CR-Set and CF-Set ($i.e.$, ${\mathbf{f}}_{ik}^{Prd}$ vs ${\mathbf{f}}_{ik}^{CF}$) highlight different areas.
(3) our DAR with the help of LR-Net and CF-Net can pay more attention to the target region than using Prd-Net alone, thus is believed to have a greater discriminative ability.

\section{Discussions}
\subsection{Hyper-parameter Settings}
\begin{figure}[h]
\centering
\includegraphics[width=1.0\columnwidth]{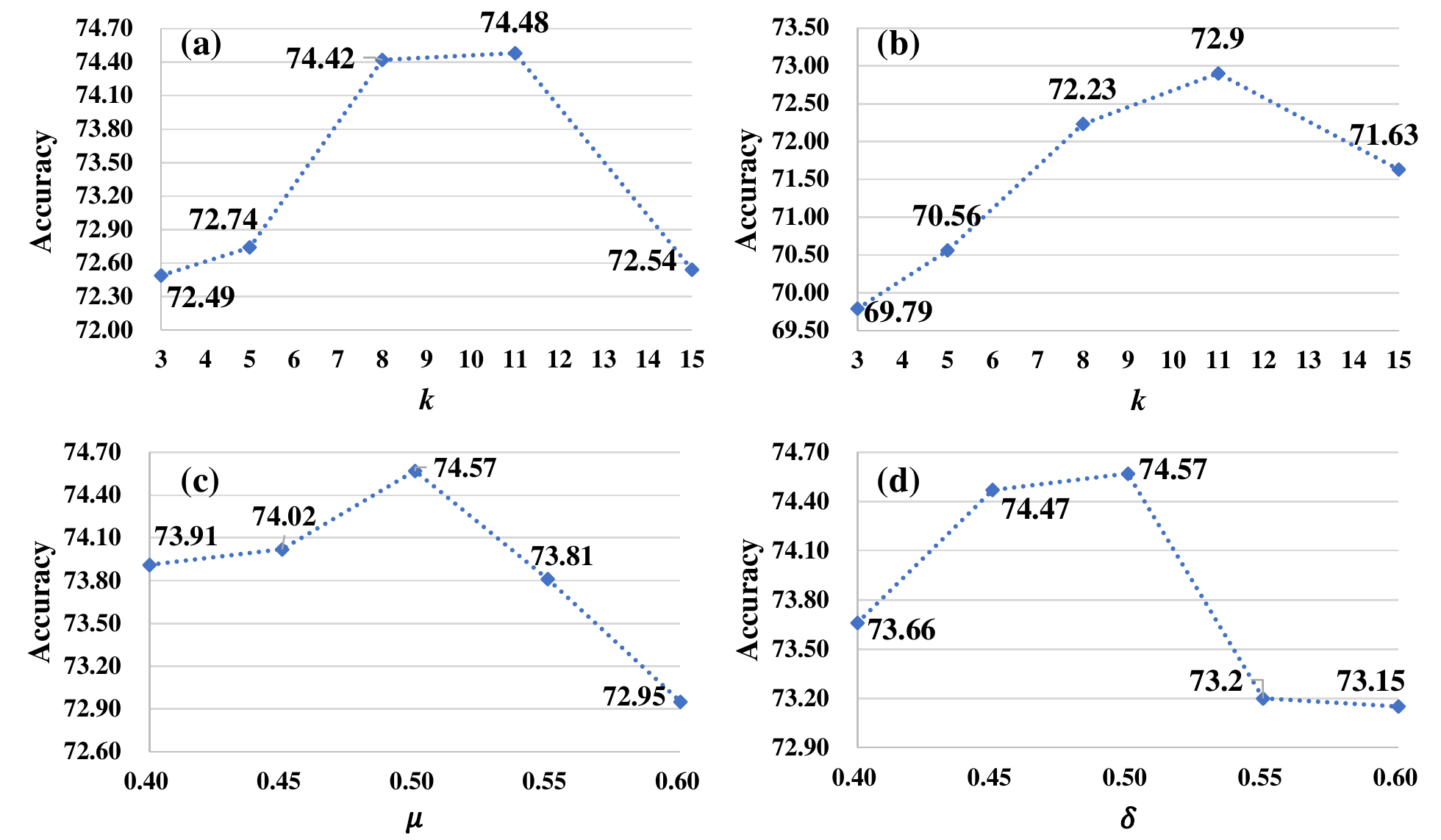}
\caption{Performance of (a) MV-Prd-CF, (b) MV-Prd-LR, and (c, d) MV-DAR when setting $k$, $\mu$ and $\delta$ to different values.}
\label{fig:fig3}
\end{figure}

In the DAR submodel, the feature maps produced by $k$-th to $m$-th convolutional blocks of LR-Net and CF-Net are transferred to the corresponding blocks of Prd-Net. The parameter $m$ is determined by the backbone and is set to 16 in our experiments. The parameter $k$ controls which blocks will be used for feature transferring. To investigate the impact of this parameter on the classification performance, we attempted to train the MV-Prd-CF model and MV-Prd-LR model with different settings of $k$ and plotted classification accuracy of both models versus $k$ in Fig.~\ref{fig:fig3}(a) and (b), respectively. It shows that both models achieve the highest accuracy when $k$ is set to 11. Thus we suggest setting $k$ to 11.

In the loss function of DAR, the weighting factors $\mu$ and $\delta$ are two critical parameters, representing the contributions of CF-Net and LR-Net, respectively. To investigate the impact of their settings on the classification performance, we compared our MV-DAR with different values of $\mu$ and $\delta$, ranging from 0.40 to 0.60 with an interval of 0.05. We plotted the accuracy versus the values of $\mu$ and $\delta$ in Fig.~\ref{fig:fig3}(c) and (d). It shows that MV-DAR achieves the highest performance when both $\mu$ and $\delta$ are set to 0.50. Hence, we empirically set $\mu$ and $\delta$ to 0.50 for this study.

\subsection{Extend to The LUNGx Challenge}
The LUNGx Challenge dataset~\cite{kirby2016lungx} in the Cancer Imaging Archive (TCIA)~\cite{Clark2013TheCI} contains 70 clinical chest CT scans with 83 lung nodules. The nodule diameters range from 3mm to 45mm. A set of 10 calibration scans was made available as a training set ($i.e.$, CR-Set). 
Five calibration scans contained a single confirmed benign nodule (two confirmed based on nodule stability for at least two years, two confirmed based on nodule resolution, and one confirmed based on a pathological assessment), and the other five contained a single pathologically-confirmed malignant nodule (two small cell carcinomas, one poorly- and one moderately-differentiated adenocarcinoma, and one non-small cell carcinoma). 
The other 60 scans with a total of 73 nodules were considered as a test set, which contained 37 benign nodules, including 13 confirmed based on nodule stability for at least two years, 19 confirmed based on nodule resolution, and five confirmed based on pathologic assessment, and 36 malignant nodules, including 15 adenocarcinomas, nine non-small cell carcinomas not otherwise specified, seven small cell carcinomas, two carcinoid tumors, one squamous cell carcinoma, and two nodules suspicious for malignancy. 

We applied our MV-DAR model, the base MV-Prd model and three ambiguous label learning methods ($i.e.$, MV-NAL~\cite{Goldberger2017TrainingDN}, MV-NLNL~\cite{Kim2019NLNLNL} and MV-D$^2$CNN~\cite{Yao2020DeepDC}) to the LUNGx Challenge dataset five times and assessed the mean and standard deviation of the AUC. 
Note that the base MV-Prd model uses the parameters trained on the LIDC-IDRI dataset for initialization. In the MV-DAR model, MV-Prd-Net is initialized with parameters fine-tuned on the LUNGx dataset ($i.e.$, the base MV-Prd model), and MV-CF-Net and MV-LR-Net are initialized with parameters trained on the LIDC-IDRI dataset.

The results in Table~\ref{tab:table4} indicate that our MV-DAR model remarkably improves AUC by 7.95\% over the MV-Prd model. Moreover, the MV-DAR model achieves the higher AUC than 12 benign-malignant nodule classification methods~\cite{Xie2019KnowledgebasedCD} and three ambiguous label learning methods, which suggests that our MV-DAR has superior performance on the LUNGx Challenge dataset.

\begin{figure}[h]
\centering
\includegraphics[width=0.75\columnwidth]{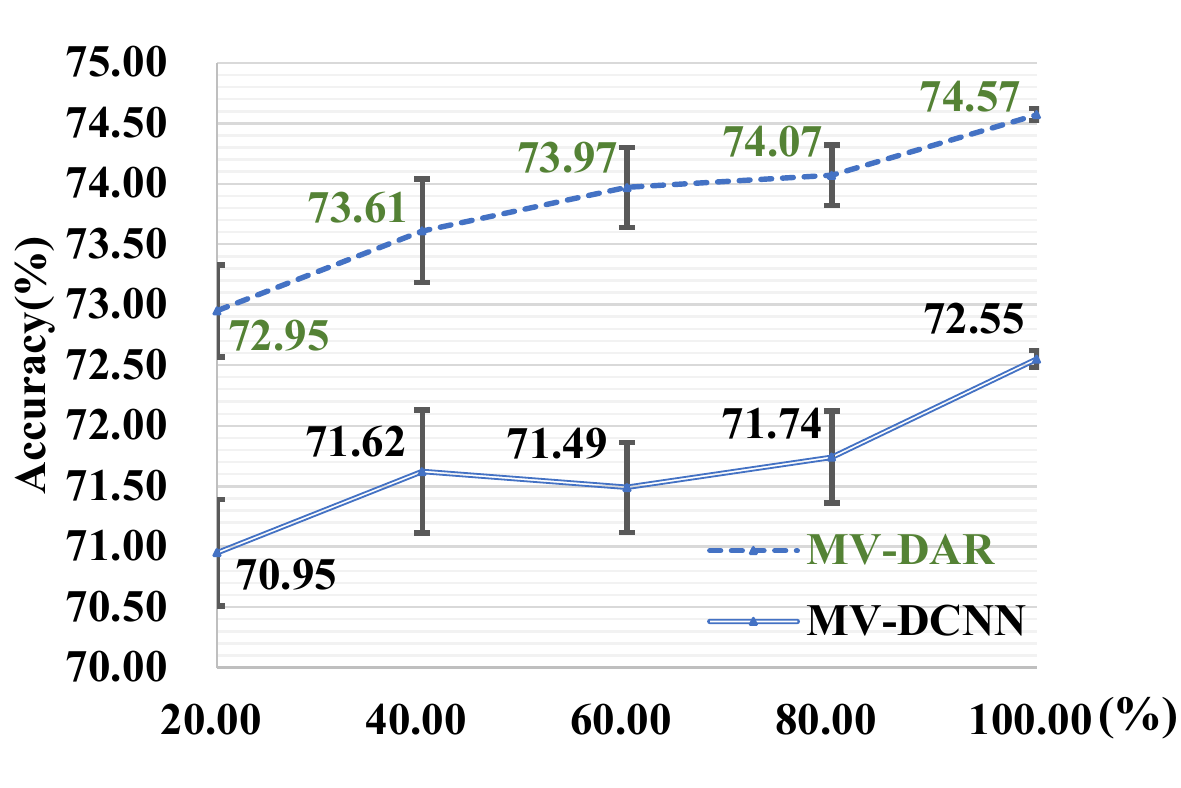}
\caption{Robustness of the MV-D$^2$CNN model and the proposed MV-DAR model to the amount of data with the ambiguous label(s).}
\label{fig:fig_noisy}
\end{figure}

\begin{figure*}[t]
\centering
\includegraphics[width=2.0\columnwidth]{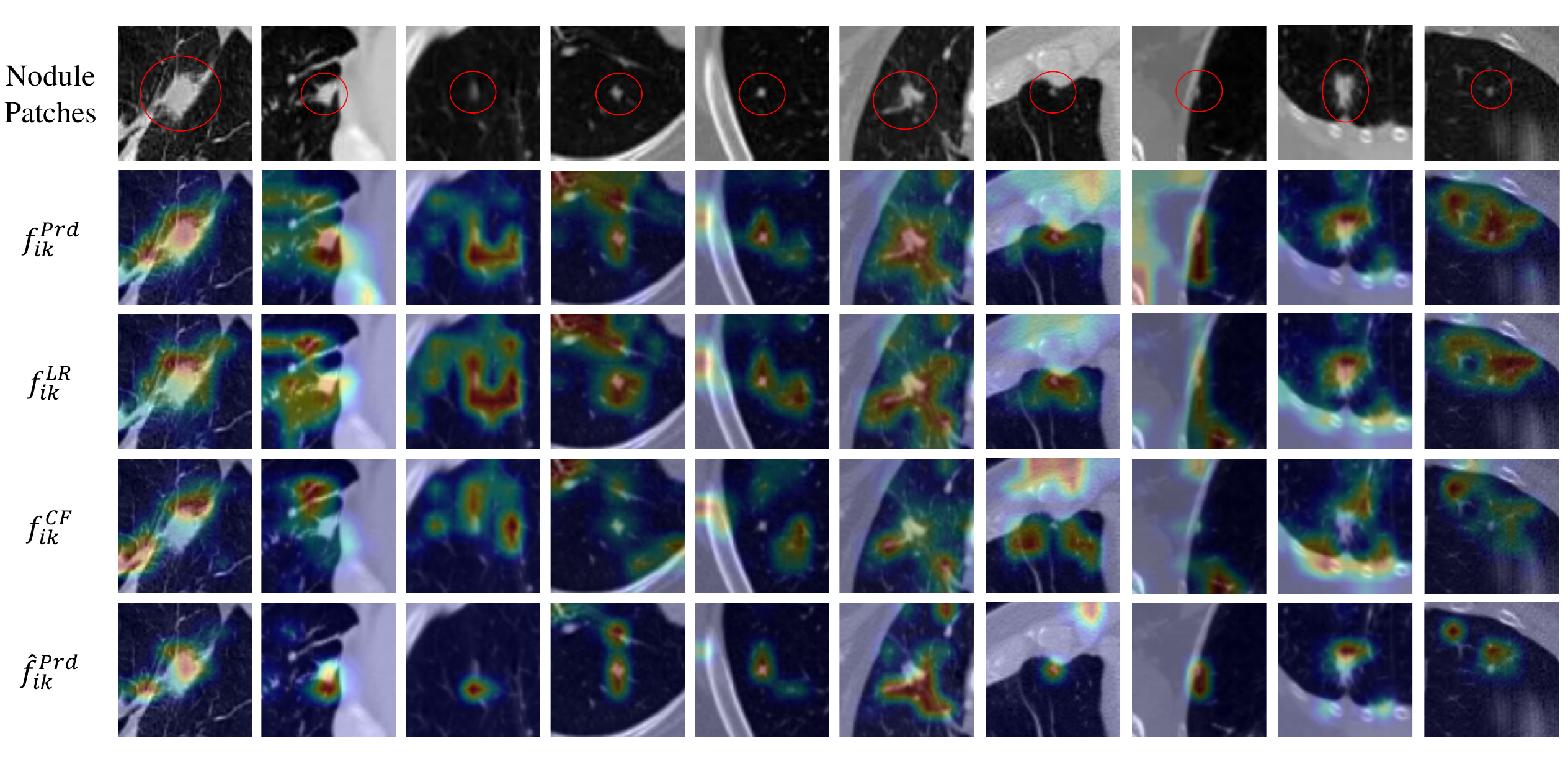}
\caption{Ten nodule patches and corresponding channel-wise sum of ${f}_{ik}^{Prd}$, ${f}_{ik}^{LR}$, ${f}_{ik}^{CF}$, and $\hat{f}_{ik}^{Prd}$.}
\label{fig:fig4}
\end{figure*}

\begin{table*}[t]
\caption{Benign-malignant lung nodule classification performance on LUNGx Challenge dataset}
\label{tab:table4}
\scriptsize
\centering
\renewcommand{\arraystretch}{1.25} 
\begin{tabular}{c|c|c|c|c}
\hline \hline
Methods & Nodule Segmentation                                                                           & Classifier                                           & AUC(\%)            & p-value   \\ \hline
1       & \multicolumn{1}{c|}{Voxel-intensity-based segmentation}                                       & SVM                                                  & 50.00$\pm$6.80     & 7.05$\times 10^{-4}$\\ \hline
2       & \multicolumn{1}{c|}{Region growing}                                                           & WEKA                                                 & 50.00$\pm$5.60     & 3.13$\times 10^{-4}$\\ \hline
3       & \multicolumn{1}{c|}{None required}                                                            & Rules based on histogram-equalized pixel frequencies & 54.00$\pm$6.70     & 1.21$\times 10^{-3}$\\ \hline
4       & \multicolumn{1}{c|}{Bidirectional region growing}                                             & Uses tumor perfusion surrogate                       & 54.00$\pm$6.60     & 1.14$\times 10^{-3}$\\ \hline
5       & \multicolumn{1}{c|}{Region growing}                                                           & WEKA                                                 & 55.00$\pm$6.70     & 1.43$\times 10^{-3}$\\ \hline
6       & \multicolumn{1}{c|}{Graph-cut-based surface detection}                                        & Random forest                                        & 56.00$\pm$5.40     & 7.00$\times 10^{-4}$\\ \hline
7       & \multicolumn{1}{c|}{Manual initialization, gray-level thresholding, morphological operations} & SVM                                                  & 59.00$\pm$6.60     & 2.78$\times 10^{-3}$\\ \hline
8       & \multicolumn{1}{c|}{None required}                                                            & Convolutional neural network                         & 59.00$\pm$5.30     & 1.15$\times 10^{-3}$\\ \hline
9       & \multicolumn{1}{c|}{GrowCut region growing with automated initial label points}               & SVM                                                  & 61.00$\pm$5.40     & 1.92$\times 10^{-3}$\\ \hline
10      & \multicolumn{1}{c|}{Radiologist-provided nodule semantic ratings}                             & Discriminant function                                & 66.00$\pm$6.30     & 1.24$\times 10^{-2}$\\ \hline
11      & \multicolumn{1}{c|}{Semi-automated thresholding}                                              & Support vector regressor                             & 68.00$\pm$6.20     & 2.15$\times 10^{-2}$\\ \hline 
-       & \multicolumn{2}{c|}{MV-KBC~\cite{Xie2019KnowledgebasedCD}}                                                                                           & 76.85$\pm$0.17     & 1.61$\times 10^{-2}$\\ \hline \hline
-       & \multicolumn{2}{c|}{MV-Prd}                                                                                                                          & 70.13$\pm$4.80     & 2.00$\times 10^{-2}$\\ \hline
-       & \multicolumn{2}{c|}{MV-NAL~\cite{Goldberger2017TrainingDN}}                                                                                          & 67.19$\pm$2.35     & 2.43$\times 10^{-4}$\\ \hline
-       & \multicolumn{2}{c|}{MV-NLNL~\cite{Kim2019NLNLNL}}                                                                                                    & 68.29$\pm$1.05     & 5.17$\times 10^{-7}$\\ \hline
-       & \multicolumn{2}{c|}{MV-D$^2$CNN~\cite{Yao2020DeepDC}}                                                                                                & 74.55$\pm$2.25     & 2.18$\times 10^{-2}$\\ \hline
-       & \multicolumn{2}{c|}{MV-DAR(Ours)}                                                                                                                    & \textbf{78.08}$\pm$0.71    & - \\ \hline \hline
\end{tabular}
\end{table*}

\subsection{Robustness to Ambiguous Labels}
To assess the impact of the number of ambiguously labeled samples on the performance of our MV-DAR model, we randomly sampled 20\%, 40\%, 60\%, or 80\%
of the data from CF-Set and LR-Set to form sub-CF-Set and sub-LR-Set, respectively.
We first pretrained MV-Prd net, MV-CF net, and MV-LR net on CR-Set, sub-CF-Set, and sub-LR-Set, respectively, and then fine-tuned MV-DAR on CR-Set.
The accuracy of MV-DAR versus the percentage of ambiguous labeled data used for training were plotted in Fig. \ref{fig:fig_noisy}.
It reveals that, with the increase of external training data, which although were ambiguously labeled, the performance of the proposed MV-DAR model has not deteriorated. On the contrary, the accuracy of MV-DAR have been improved monotonously. 
It proves that our MV-DAR model is robust to the ambiguous label(s).

Moreover, we also verified the robustness of the MV-D$^2$CNN model, the most competitive model in Table~\ref{tab:table2}, to ambiguous labels. It can be seen from Fig. \ref{fig:fig_noisy} that the model gain of MV-DAR is greater than that of MV-D$^2$CNN in accuracy, suggesting that our MV-DAR model has a stronger ability to learn discriminative information from ambiguous labels than MV-D$^2$CNN.

\subsection{Computational Complexity}
The proposed MV-DAR was implemented using Pytorch software packages and was evaluated using a desktop with an NVIDIA Tesla P100 GPU. In the pretraining stage, it took about 0.5 hours, 2.5 hours, and 12 hours to train Prd-Net, LR-Net, and CF-Net, respectively. In the fine-tuning stage, it took about an hour to train the entire model. When testing, it costs less than 0.15 seconds to diagnose each nodule on average. The fast inference suggests that our MV-DAR could be used in a routine clinical workflow.

\section{Conclusion}
We propose the MV-DAR model to learn from ambiguous labels, including inconsistent malignancy scores given by multiple radiologists and an unreliable single score given by only one radiologist, for lung nodule malignancy prediction.
Our results on the LIDC-IDRI dataset indicate that the proposed MV-DAR model is substantially superior to four proxy-label based methods and an identification-based method.
The verification on the LUNGx dataset shows that our MV-DAR model can be successfully generalized to this small dataset.
Meanwhile, the ablation studies not only demonstrate the effectiveness of each part of MV-DAR, but also confirm its ability to learn the image representation ability from ambiguous labels ($i.e.$ the performance improves monotonously with the increase of ambiguously labeled training data).
Although building upon the application of lung nodule malignancy prediction, our MV-DAR model itself is generic and can be extended to other tasks troubled by ambiguous labels. In our future work, we plan to investigate the incorporation of the annotation confidence estimated based on other data modalities such as clinical genomics into the diagnosis model.

\section*{Acknowledgment}

We acknowledge the National Cancer Institute (NCI) and the Foundation for the National Institutes of Health, and their critical role in the creation of the free publicly available LIDC-IDRI dataset used for this study.
We also appreciate the efforts devoted by the organizers and sponsors of the SPIE-AAPM-NCI Lung Nodule Classification (LUNGx) Challenge to collect and share the data for comparing automated nodule classification algorithms.

\end{document}